\documentclass[aps,prb,twocolumn,superscriptaddress,
preprintnumbers,amsmath,amssymb,floatfix]{revtex4}
\usepackage{graphicx, epstopdf}
\usepackage{dcolumn}
\usepackage{bm}

\begin{document}


\title{Transforming Fabry-P\'erot resonances into a Tamm mode}


\author{Maxim Durach}
\email{mdurach@georgiasouthern.edu}
\affiliation{Physics Department, Georgia Southern University, Statesboro, GA 30460-8031}
\author{Anastasia Rusina}
\affiliation{Physics Department, Georgia Southern University, Statesboro, GA 30460-8031}


\date{\today}

\begin{abstract}

We propose a novel photonic structure composed of
metal nanolayer, Bragg mirror and metal nanolayer. 
The structure supports resonances that are transitional between Fabry-P\'erot and Tamm modes.
When the dielectric contrast of the DBR is removed 
these modes are a pair of conventional Fabry-P\'erot resonances.
They spectrally merge into a Tamm mode at high contrast.
Such behavior differs from the results for structures supporting Tamm modes
reported earlier. The optical properties of the structure in the frequency range of the DBR stop band, including highly beneficial $50\%$ transmittivity through thick structures,
are determined by the introduced in the paper hybrid resonances.
The results can find a wide range of photonic applications.

\end{abstract}

\pacs{42.70.Qs, 78.55.Cr, 41.20.Jb, 42.25.Hz, 78.68.+m, 73.20.Mf}
\keywords{Tamm states, Bragg mirror, thin films, planar structures, Fabry-Perot resonances, optoelectronics}

\maketitle

Metal inclusions are increasingly employed to improve the photonic performance
of semiconductor devices. 
For example, resonant-cavity enhancement by thin metal layers 
has been shown to increase the quantum efficiency 
of novel multi-band photodetectors \cite{Perera_resonant_cavity_photodetectors}.  
The interface between metal and distributed Bragg reflector (DBR),
a configuration often encountered in optoelectronic devices,
has attracted a great deal of attention since the seminal papers proposing
and observing optical Tamm states \cite{Kavokin_Shelykh_optical_Tamm_modes} 
and Tamm plasmons (TP) \cite{Kaliteevski_TP_prediction, Sasin_et_al_Tamm_plasmon_observation}.
Several developments soon followed, such as studies of hybrid 
optical Tamm states \cite{Bruckner_hybrid_optical_Tamm_states}
and coupled TPs \cite{Iorsh_et_al_coupled_Tamm_plasmons}. 
Interaction of Tamm states with excitons \cite{Symonds_Tamm_plasmon_exciton_coupling}
have been observed and Tamm exciton polaritons
are proposed as candidates for signal carriers in a novel type of integrated circuits \cite{Liew_at_al_Tamm_plasmon_exciton_integrated_circuits, Kaliteevski_et_al_channeling_Tamm_plasmons}.
TPs have been found promising for the fields of quantum optics
\cite{Gazzano_et_al_Tamm_plasmons_under_disks_quantum_emitter}
and polaritonics \cite{Flayac_et_al_vortices_in_polariton_superfluids}.

In this paper we study optical properties of the
metal nanolayer - DBR - metal nanolayer (MNL-DBR-MNL) structure 
in the spectral region of the stop band of the DBR for the first time.
It is shown that the proposed structure supports modes that are intermediate between
Fabry-P\'erot resonances (FPRs) and a Tamm plasmon.
The dependence of the modes on the dielectric contrast of the DBR
is surprising in light of previous reports (e.g.  Ref. [\onlinecite{Iorsh_et_al_coupled_Tamm_plasmons}]), 
but is very reasonable according to the given in our paper arguments.
Another distinctive feature of the structure is the symmetry of the resonances.
We demonstrate that magnetically symmetric resonance (i.e. electrically antisymmetric)
is at low energy.
The intriguing properties of the structure are achieved if the 
metal layers are thinner than metal skin depth, i.e. $\lesssim 25~\mbox{nm}$.
Then at resonances the incident power is equally split between
absorption and transmission. Transmission of $50\%$ of incident power persists until
the resonances merge into a TP.
Owing to the sensitivity of the modes to the thickness of the metal layers,
dielectric contrast of the DBR and to the number of layers, 
the modes can be used to tailor response 
of planar metal-semiconductor devices and to control field distribution inside them. 

\begin{figure}
\includegraphics[width=.5\textwidth]{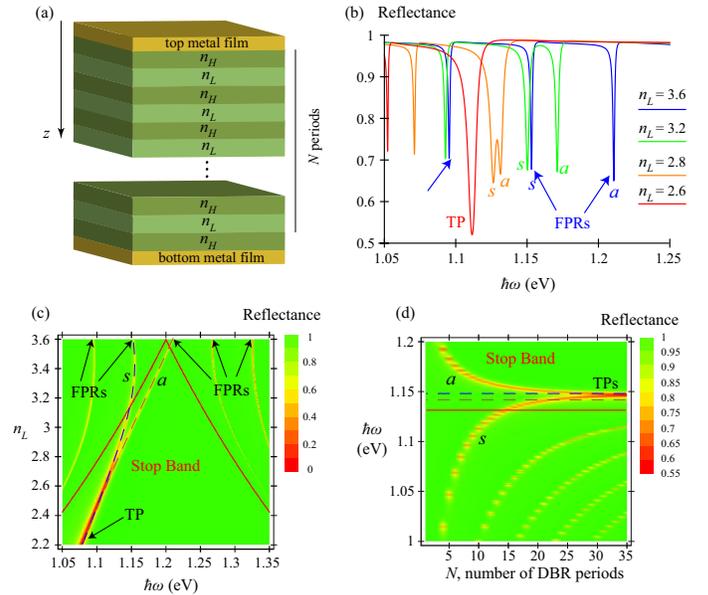}
\caption{(a) The metal nanolayer - distributed Bragg reflector - metal nanolayer (MNL-DBR-MNL) structure.
(b) Reflectance spectrum for different dielectric contrasts. Metal layer thickness $d_m = 50~\rm{nm}$, $N = 20$. 
(c) Reflectance spectrum dependence on the dielectric contrast. Metal layer thickness $d_m = 50~\rm{nm}$, $N = 20$. 
(d) Dependence of the reflectance spectrum on $N$ for $d_m = 50~\rm{nm}$ and $n_L = 3$. 
The reflectance is color coded to the right of the graphs.}
\label{Reflectivity_thick_metal.eps}
\end{figure}

The proposed structure is composed of two identical metal films and 
$(HL)^N H$ one-dimensional array, which forms a DBR with Bragg frequency $\omega_B$
(see Fig.\ \ref{Reflectivity_thick_metal.eps} (a)).
Each of $N$ periods of the DBR is composed of a couple of dielectric layers with refraction indices $n_H$ and $n_L$, $n_H>n_L$.
The thicknesses of the layers are denoted as $d_H=\pi c/(2 n_H \omega_B)$ and 
$d_L=\pi c/(2 n_L \omega_B)$, making the total thickness of the DBR stack $D = (N+1) d_H + N d_L$.
The thickness of metal films covering the DBR is $d_m$.

We follow the characteristic matrix approach \cite{Born_Wolf}.
The tangential components $H(z)$ and $E(z)$ of magnetic and electric field
in a planar structure can be related at locations $z_0$ and $z$ through the following matrix equation
\begin{equation}
 \left(
     \begin{array}{c}
      H(z_0) \\
      E(z_0)
     \end{array} \right) = M  
\left(
     \begin{array}{c}
      H(z) \\
      E(z)
     \end{array} \right)~.~~
\label{tmm}
\end{equation}
Using the Abeles theorem 
\cite{Abeles_matrix_power} we have obtained an expression for
the characteristic matrix of the $(HL)^N H$ array
\begin{equation}
M_{DBR} = M_H U_N(a)+ M_L^{-1} U_{N-1}(a)~,~~
\label{DBR_tm}
\end{equation}
where $M_H$ and $M_L$ are the characteristic matrices of the $H$ and $L$ layers,
$U_N$ are the Chebyshev polynomials of the 2nd kind, $a=\cos(k_B d)$,
were $k_B$ and $d = d_H+d_L$ are the Bloch vector and the thickness of a DBR period.

The transmission spectrum of the DBR is composed of 
bands of strong transmission alternating with stop bands, in which the transmission is low. 
In the stop bands the Bloch vector $k_B$ is complex and fields decay
from the edges of the DBR.
The decaying fields in the DBR can be matched at the metal-DBR boundary
with fields penetrating the metal and form the TP modes whose intensity is enhanced at the boundary
\cite{Kaliteevski_TP_prediction}.
We consider a structure in which the DBR is bounded by metal layers from both sides.
If the metal layers are thicker than the metal skin depth $l_s$, i.e.
$d_m > l_s=(k_0 \rm{Re}\sqrt{-\varepsilon_m})^{-1}\approx 25~\rm{nm}$,
the structure supports modes with the following dispersion
\begin{equation}
 \left(
     \begin{array}{c}
      1 \\
      - p_m
     \end{array}
   \right) =\pm M_{DBR}
   \left(     
   \begin{array}{c}
      1 \\
       p_m
 \end{array}
   \right)~,~~
\label{thick_M_DBR_M}
\end{equation}
or $p_m = (\pm 1-m_{11})/m_{12} = -m_{21}/(\pm1+m_{11})$, 
where $m_{ik}=(M_{DBR})_{ik}$. 
In the paper we use a notation $p$ to denote $k/(k_0\varepsilon)$
with subscripts $i$, $m$, $H$, $L$ and $s$ corresponding 
to the medium of incidence, metal films, $H$ and $L$ layers and substrate.
Here $k_0=\omega/c$ 
and $\varepsilon$ is the dielectric permittivity of a medium. The structures we consider 
are non-magnetic.
In Eq.\ (\ref{thick_M_DBR_M}) the plus sign is for modes
whose magnetic field is symmetric with respect to the center of the structure,
and minus for antisymmetric modes.
In absence of dielectric contrast in the DBR, $n_L=n_H$, the modes of Eq.\ (\ref{thick_M_DBR_M}) turn into
guided modes of the resulting metal-insulator-metal (MIM) structure.

Increasing the dielectric contrast of the DBR expands the stop bands and 
induces frequency shifts of pairs of the guided modes 
closest to the stop bands toward each other. 
This effect can be seen from Eqs.\ (\ref{DBR_tm}) and (\ref{thick_M_DBR_M}). 
The Bloch wave vector at frequencies close to the first stop band can be represented as
$k_B = (\pi+\delta)/d$, with $\delta$ being imaginary within the stop band.
From the definition of Chebyshev polynomials $U_N(\cos(x)) = \sin((N+1)x)/\sin(x)$
in the limit $ N \delta \gg 1$  (i.e. when field strongly decays on the length of the DBR) 
it follows that the matrix elements of $M_{DBR}$ exponentially grow such that
the symmetric and antisymmetric modes of Eq.~(\ref{thick_M_DBR_M}) become degenerate 
satisfying $p_m = - m_{11}/m_{12}$. 
The dispersion relation for this degenerate mode is
\begin{equation}
p_m =- i p_H \cot(k_H d_H)\frac{2 \cosh(\delta/2)}{e^{\delta/2}-(p_H/p_L)~ e^{-\delta/2}}~,~~
\label{TP_thick_M_DBR_M}
\end{equation}
where $k_H$ is the $z$ component of the wave vector in $H$ layer.
Excluding $\delta$ from Eq.\ (\ref{TP_thick_M_DBR_M})
and definition of the Bloch vector 
in the range $|\omega-\omega_B|\ll \omega_B$ we arrive at 
\begin{equation}
\frac{\omega-\omega_B}{\omega_B} 
\approx -\frac{2 i (n_H-n_L)}{\pi n_m}\frac{n_H n_L}{(n_H^2-n_H n_L+n_L^2)}~~
\label{TP_thick_M_DBR_M_apx}
\end{equation}
for normal incidence. With the exception of the last factor $\approx 1$,
Eq.\ (\ref{TP_thick_M_DBR_M_apx})  
is similar to the original expression of Ref.\ [\onlinecite{Kaliteevski_TP_prediction}] for the TP frequency.
The main approximation of Ref.\ [\onlinecite{Kaliteevski_TP_prediction}]
turns out to be the assumption $|n_m|\gg \sqrt{n_H n_L}$, resulting in error of
$\approx 10~\mbox{meV}$ (see Fig.\ \ref{Reflectivity_thick_metal.eps}(d)). 

Thus, as the dielectric contrast of the DBR is increased
in a MNL-DBR-MNL structure with thick metal films
a pair of guided modes near Bragg frequency gradually become degenerate and 
are transformed into a TP. The splitting of TPs
in a DBR bounded by two cavities with thicknesses of $d=d_H-l_s$ and semi-infinite metal layers
was considered in Ref.\ [\onlinecite{Iorsh_et_al_coupled_Tamm_plasmons}]. 
The splitting for that structure grows with reduction of dielectric contrast, 
but then collapses at moderate contrasts according to results of
Ref.\ [\onlinecite{Iorsh_et_al_coupled_Tamm_plasmons}],
which is a very different behavior from the resonances considered in our structure.

The optical properties of the MNL-DBR-MNL structure are described by the equation:
\begin{equation}
 \left(
     \begin{array}{c}
       (1+r) \\
       p_0 (1 - r)
     \end{array}
   \right)=M_m M_{DBR} M_m  
   \left(     
   \begin{array}{c}
       t \\
       p_s t
     \end{array}
   \right)~,~~
\label{M_DBR_M_equation}
\end{equation}
where $r$ and $t$ are the reflectance and transmittance amplitudes and $M_m$ is the characteristic matrix of the metal films.
From Eq.\ (\ref{M_DBR_M_equation}) using general relations between the
characteristic and transfer matrices \cite{Furman_Tikhonravov} we have obtained the analytical expressions for $r$ and $t$. 
For instance, the transmittance amplitude for normal incidence is given by
\begin{equation}
t =\frac{t_{im} t_{Hm} e^{i(2 N \beta + \alpha)}}{1-r_{Hm}^2 e^{2 i(2 N \beta  + \alpha)}
+e^{i(2 N\beta + \alpha)}\sin(2 N \beta)\gamma
}
~,~~
\label{transmission_amplitude}
\end{equation}
where  
$t_{im}=2 p_i/(p_m+p_i)$, $t_{Hm}=2 p_H/(p_m+p_H)$ and $r_{Hm}=(p_H-p_m)/(p_m+p_H)$
represent the Fresnel formulae, while $\alpha=\pi\omega/(2 \omega_B)$, $\beta= k_B d/2$. 
We also use shorthand notations for 
$\gamma=\sec(\beta)(\zeta\sin(\beta-\alpha)- \xi \csc(\beta)\sin(\alpha))$, 
$\zeta=(-1+r_{Hm}^2-i(1+r_{Hm}^2)\cot(\alpha))$ and 
$\xi =i r_{Hm} (p_H^2-p_L^2)/(2 p_H p_L)$. 
The analytical expressions were used to check the results of the numerical computations 
obtained by 
solution of Eq.\ (\ref{M_DBR_M_equation}). 

In Figs.\ \ref{Reflectivity_thick_metal.eps}(b)-(d)  we show computation of reflectance 
of a MNL-DBR-MNL structure under normal incidence. 
The medium of incidence and the substrate are considered as vacuum $p_0=p_s=1$.
The phenomenological parameters of gold are used for metal films \cite{Johnson_Christy}. 
Their thickness is taken $d_m = 50~\rm{nm}$
which exceeds the skin depth $l_s\approx 25~\mbox{nm}$ by factor of $2$. The DBR is formed by 
GaAs/GaAlAs structure. The Bragg frequency is selected to be $\omega_B = 1.2~\mbox{eV}$.
In this frequency range index of refraction for GaAs is $n_H = 3.6$,
while the index of refraction of GaAlAs component physically can be in the
range $n_L=2.9-3.6$ for different Al fraction \cite{Pikhtin_AlGaAs_optical_constants}. 
For the sake of argument we vary $n_L$ from $2$ to $3.6$ in this paper. 

In absence of the dielectric contrast $\beta=\alpha$, $\xi=0$ and Eq.\ (\ref{transmission_amplitude})
transforms into the 
expression for transmission through a Fabry-P\'erot resonator \cite{Born_Wolf}.
In MNL-DBR-MNL structures with low dielectric contrast the transmittance is high 
(and reflectance is low) at Fabry-P\'erot resonances, 
while all the incident radiation is reflected at other frequencies. 
In Fig.\ \ref{Reflectivity_thick_metal.eps}(b) we show the reflectance spectrum $R = |r|^2$.
The structure for which $n_L=n_H=3.6$ (blue curve) has three FPRs 
in the frequency range shown (marked by arrows) separated by equal frequency bands 
$\Delta \omega \approx 60 \rm{meV}$ of strong reflection. 
If the index of refraction is lowered to $n_L=3.2$ (green curve) one of the FPRs corresponding to 
a magnetically antisymmetric mode (marked by $a$) is strongly shifted to the red.
The symmetric mode at lower frequency (marked by $s$) 
is not shifted, which leads to reduced separation 
$\Delta \omega \approx 20 \rm{meV}$ between the modes.
At $n_L = 2.8$ (orange curve) the reflection dips 
corresponding to the symmetric and antisymmetric modes are separated by 
$\Delta \omega \approx 5 \rm{meV}$. The energies of the modes fully merge at $n_L=2.6$ 
as shown by the red curve and correspond to the energy of TP.

The transformation of the modes is further demonstrated in Fig.\ \ref{Reflectivity_thick_metal.eps}(c).
This panel shows the dependence of the reflectance spectrum on the dielectric contrast (i.e. on $n_L$).
The stop band lower and upper frequencies are shown by the red curves 
(they formally limit the frequency band with $|\mbox{Im}~k_B| \neq 0$).
At high dielectric contrast the MNL-DBR-MNL structure fully reflect
the incident radiation except for the narrow band marked as TP. 
As the dielectric contrast is reduced this reflection dip is split into two dips,
which become FPRs at low contrast.
We show the dispersion of the modes [Eq.\ (\ref{thick_M_DBR_M})] by the dashed lines.
They follow the reflectance dips obtained using Eq.\ (\ref{M_DBR_M_equation}).
The correspondance of the reflection dips to the modes of Eq.\ (\ref{thick_M_DBR_M}) is 
due to the fact that $d_m$ considerably exceeds the skin depth $l_s$.
At high dielectric contrast the modes described by Eq.\ (\ref{thick_M_DBR_M}) merge and
follow the dispersion of the TP given by Eq.\ (\ref{TP_thick_M_DBR_M}).

The Tamm plasmon splitting as the contrast of the DBR is reduced is due to reduction of  $|\mbox{Im}~k_B|$
and hybridization of TPs localized at different metal-DBR boundaries. 
Along the same lines the splitting should be observed if the number of periods $N$ of the DBR is reduced
at constant contrast (i.e. $|\mbox{Im}~k_B| = const$). This is confirmed by the 
Fig.\ \ref{Reflectivity_thick_metal.eps}(d).
If $N>25$ one can see a single reflectance dip at the bottom of the stop band
due to the excitation of the TP. The dashed blue line corresponds to the energy of TP
calculated using Eq.\ (\ref{TP_thick_M_DBR_M}). The dashed brown line is at the energy of TP
calculated using expression of Ref. [\onlinecite{Kaliteevski_TP_prediction}]
(the discrepancy of about $10~\mbox{meV}$). For $N<25$ this dip is split into two dips with higher frequency
separation between them for smaller $N$, reaching $150~\mbox{meV}$ for the structure with $N=5$.

\begin{figure}
\includegraphics[width=.5\textwidth]{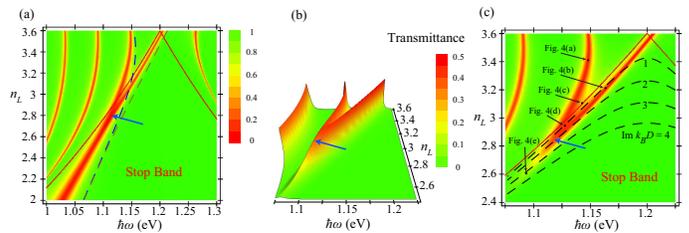}
\caption{Optical properties of MNL-DBR-MNL structure with $d_m = 20~\rm{nm}$ below the skin depth. 
In all panels number of periods $N = 20$. (a) Dependence of the reflectance spectrum on dielectric contrast.
(b)-(c) The same for transmittance. 
}
\label{Reflectivity_thin_metal.eps}
\end{figure}

If the thickness of the metal layers is smaller than $l_s$, the frequencies of
the reflection dips cannot be described by  Eqs.\ (\ref{thick_M_DBR_M}) and (\ref{TP_thick_M_DBR_M}).
In Fig.\ \ref{Reflectivity_thin_metal.eps}(a) we show computations similar 
to those shown in Fig.\ \ref{Reflectivity_thick_metal.eps}(c),
but for $d_m = 20~\mbox{nm}$. 
A considerable shift to the red of both resonances compared to the modes 
in the structure with thick metal films (dashed lines) is observed. 
Reflectance reduction at the resonances is much stronger than in the case with $d_m=50~\mbox{nm}$.
This is due to the enhanced interaction of incident radiation with the resonances for $d_m\lesssim l_s$.
The resonances are strongly excited leading to higher field intensity in
the structure and increased transmittance.

In  Figs.\ \ref{Reflectivity_thin_metal.eps}(b)-(c) we plot the transmittance $T=p_i |t|^2/p_s$ of the same MNL-DBR-MNL
structure as in Fig.\ \ref{Reflectivity_thin_metal.eps}(a).
The transmittance is highly suppressed in the frequency band under consideration, 
except for the strong transmittance peaks at the resonances, corresponding to the dips in reflectance.
It can be seen that at lower contrast, when the transmittance peaks are split,
the maximum transmittance is $T \approx 0.5$, while reflectance is $R\approx 0$, which 
allows to estimate the absorbance by both modes as $A = 1- R- T \approx 0.5$. 

The resonances merge if the dielectric contrast of the DBR is increased. The merger point, 
which corresponds to $n_L \approx 2.8$, 
is marked by the blue arrows in all panels of Fig.\ \ref{Reflectivity_thin_metal.eps}. 
For even stronger dielectric contrast the transmittance decays exponentially as can be clearly seen
from Fig.\ \ref{Reflectivity_thin_metal.eps}(b). This decay can be understood
from Fig.\ \ref{Reflectivity_thin_metal.eps}(c), in which the transmittance is superimposed
on the lines of constant $\mbox{Im}~k_B D$ (the dashed black lines). 
As the contrast is increased the frequency of the modes correspond to higher value of this
parameter, for instance, $\mbox{Im}~k_B D\approx 4$ for $n_L=2.5$. 
From Eq.\ (\ref{transmission_amplitude}) one can see that for higher contrast
$t\propto \exp(-\mbox{Im}~ k_B D)$. This implies that optical
fields become exponentially weak on the bottom side of the 
structure 
with consequent diminishing of transmittance. We will show below 
that decay of transmittance is due to
destructive interference between the resonant modes on the bottom side of the DBR.

Consider a structure with inversion symmetry such that $p_s = p_0$. The tangential magnetic and electric field
components can be represented as sums of even and odd functions with respect to the center of the structure, 
e.g. as $H(z^\prime) = (H(z^\prime)+H(-z^\prime))/2+(H(z^\prime)-H(-z^\prime))/2$,
where $z^\prime=z-D/2$.
Both symmetric and antisymmetric terms represent solutions of Maxwell equations for the structure. 
Therefore, Eq.~(\ref{M_DBR_M_equation}) is partitioned as:
\begin{equation}
\left(
    \begin{array}{c}
     (1+r_{s,a})\\
       p_0 (1 - r_{s,a})
    \end{array}
   \right)=\pm M  
   \left(     
   \begin{array}{c}
       (r_{s,a} + 1) \\
      p_0 (r_{s,a} -1)    
   \end{array}
   \right)~,~~
  \label{M_DBR_M_separation}
\end{equation}
where index $s$ ($a$) corresponds to plus (minus) sign,
so that $r=(r_s+r_a)/2$ and $t=(r_s-r_a)/2$. In Eq.\ (\ref{M_DBR_M_separation}) $M=M_m M_{DBR} M_m$.
All of the spectral characteristics of the structure with inversion symmetry
are determined by the reflection amplitudes $r_s$ and $r_a$ 
for the magnetically symmetric and antisymmetric parts of the field.

\begin{figure}
\includegraphics[width=.5\textwidth]{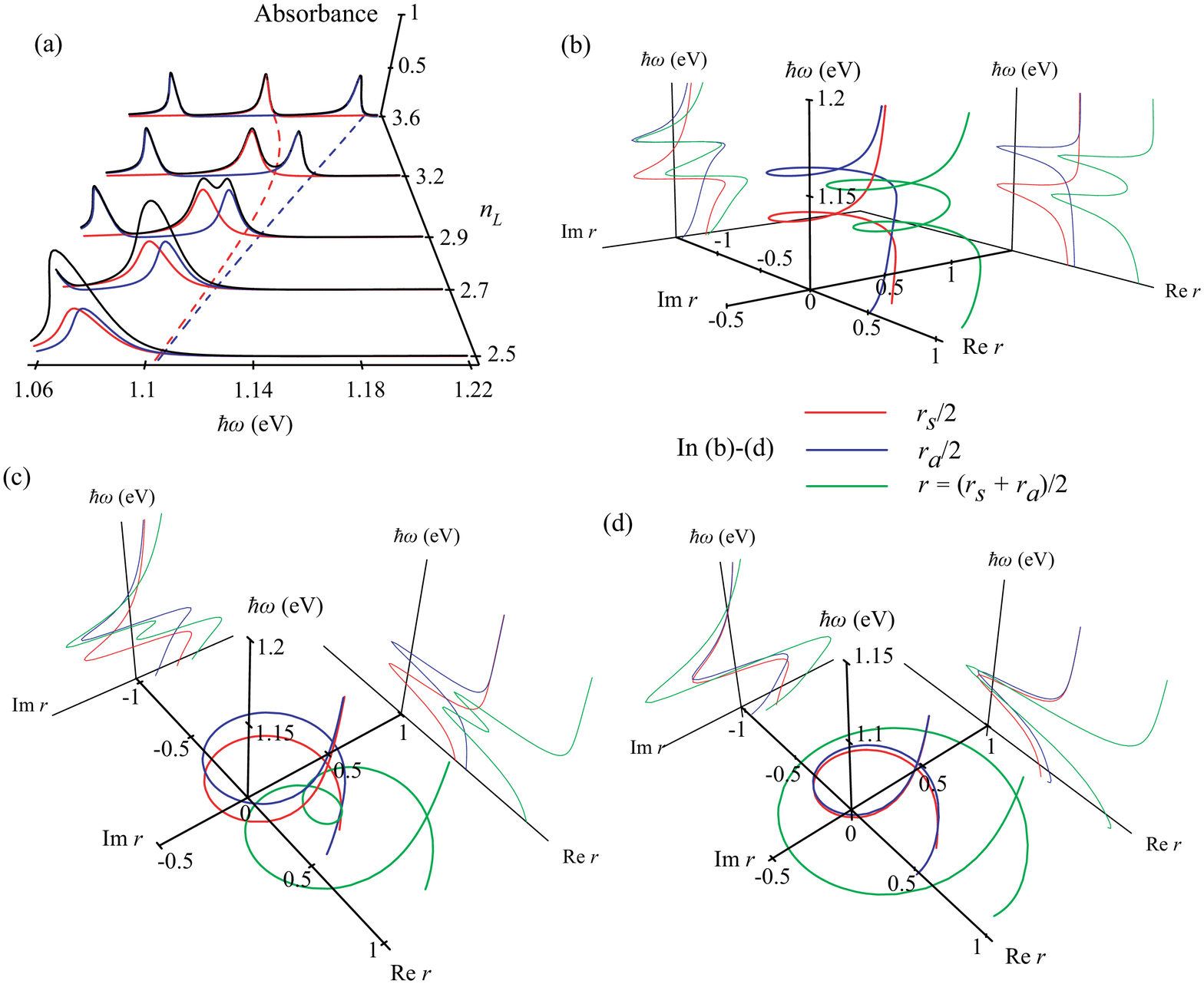}
\caption{(a) Absorbance by the 
structure with $d_m = 20~\rm{nm}$ and $N = 20$.
Red and blue curves are the partial absorbances of symmetric and antisymmetric field components.
The black curves represent the full absorbance.
(b)-(d) The reflectance coefficients as complex functions of frequency.
The projections of the 3-dimensional curves are shown on the sides to ease the perception. 
The dielectric contrast is varied from panel to panel:
(b) $n_L=3.2$ (c) $n_L=3.$ (d) $n_L=2.5$.}
\label{Absorbance_modes.eps}
\end{figure}

The absorbance spectra of the same structure as in Fig.\ \ref{Reflectivity_thin_metal.eps}
are shown in Fig.\ \ref{Absorbance_modes.eps}(a).
In Fig.~\ref{Absorbance_modes.eps}(a) we plot the total absorbance $A=1-|r_s|^2/2-|r_a|^2/2$ (black curves)
as well as partial absorbances of the symmetric $A_s=(1-|r_s|^2)/2$ (red curves) and 
antisymmetric $A_a=(1-|r_a|^2)/2$ (blue curves) field components.
The absorbance peaks in Fig.\ \ref{Absorbance_modes.eps}(a), which are in agreement with  
reflectance dips and transmittance peaks of Fig.~\ref{Reflectivity_thin_metal.eps}.
Using the partial absorbances and coefficients $r_s$ and $r_a$ the lower energy peak
can be attributed to the resonance of the magnetically symmetric part of the field,
while the higher energy peak is related to the antisymmetric part of the field. 
As was mentioned above
at low dielectric contrast the absorbance for both modes is $A\approx 0.5$.
When the peaks of partial absorbances merge the total absorbance increases, 
which corresponds to the exponential decay of the transmittance (cf. Figs.\ \ref{Reflectivity_thin_metal.eps}(b)-(c)).

In  Fig.\ \ref{Absorbance_modes.eps}(b)-(d) we show the frequency dependences of complex parameters
$r_s/2$ (red curve), $r_a/2$ (blue curve) and the total reflectance amplitude $r=(r_s+r_a)/2$ (green curves).
In panel (b) the structure with moderate dielectric contrast ($n_L = 3.2$) is considered.
At lower frequencies both parts of the field are non-resonant, so that $r_s/2\approx r_a/2\approx 0.5$
and all of the incident power is reflected for both field symmetries.
At $\hbar \omega=1.14~\mbox{eV}$ the symmetric part of the field becomes resonant, which
can be seen by the loop of the red curve in Fig.\ \ref{Absorbance_modes.eps}(b). 
At $\hbar \omega=1.163~\mbox{eV}$ the antisymmetric part of the field is resonant 
and the blue curve makes a resonance loop. Both loops are well separated in frequency, 
which results in two separate loops of $r$, two dips of reflectance $R=|r|^2$ 
(cf. Fig.\ \ref{Reflectivity_thin_metal.eps}(a)) and two
peaks of transmittance $T=|r_s-r_a|^2/4$ (cf. Figs.\ \ref{Reflectivity_thin_metal.eps}(b)-(c)) 
and absorbance $A$ (cf. Fig.\ \ref{Absorbance_modes.eps}(a)). 
In the midst of the resonances the reflection coefficients 
$r_s\approx r_a\approx -0.3$  are phase-shifted by $\pi$ with respect to the incident waves.
The phase of the total reflection coefficient $r$ varies within the range from $-\pi/4$ to $\pi/4$.

\begin{figure}
\includegraphics[width=.5\textwidth]{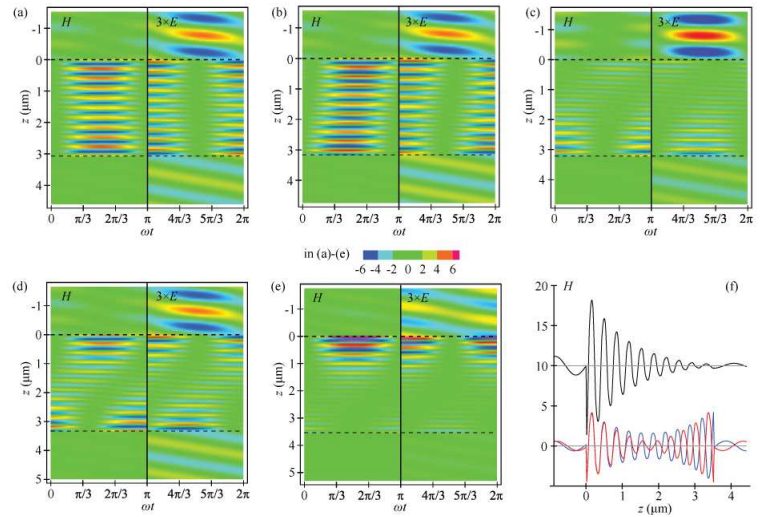}
\caption{Optical field distribution during an optical period.
The magnetic field $H(z,t)$ is shown during the first half of the period,
the electric field $E(z,t)$ multiplied by 3 is shown during the second half.
In panels (a)-(e) the incident field amplitude is unity (see the color scale between the panels).
The dielectric contrast and frequencies are selected as follows:
(a) $n_L = 3.4$ and $\hbar \omega = 1.147~\mbox{eV}$,
(b) $n_L = 3.2$ and $\hbar \omega = 1.163~\mbox{eV}$,
(c) $n_L = 3.1$ and $\hbar \omega = 1.144~\mbox{eV}$,
(d) $n_L = 2.9$ and $\hbar \omega = 1.123~\mbox{eV}$,
(e) $n_L = 2.6$ and $\hbar \omega = 1.093~\mbox{eV}$.
(f) Magnetic field for the same parameters as in (e) 
shown at $t=\pi/(2\omega)$. Red and blue curves represent 
the symmetric and antisymmetric components of the field. Black curve shows the 
total field. The black curve is offset by 10 for clarity.}
\label{field_graphs.eps}
\end{figure}

Increasing the contrast (see Fig.\ \ref{Absorbance_modes.eps}(c) for $n_L = 3$ 
and Fig.\ \ref{Absorbance_modes.eps}(d) for $n_L = 2.5$) results in reduction 
of splitting between symmetric and antisymmetric resonances. Correspondingly, the loops that $r_s$ 
and $r_a$ make in the complex plane as functions of frequency occur at closer frequencies,
which changes the shape of the curve corresponding to the total reflectance coefficient $r$. 
Instead of the two frequency-separated loops,  
$r$ makes a heart-shaped loop at $n_L = 3$ with a small sub-loop that corresponds to the increase 
in reflectance between closely placed dips in the spectrum shown in Fig.\ \ref{Reflectivity_thin_metal.eps}(a).
At $n_L=2.5$ the two resonances merge and the loops coincide in
Fig.\ \ref{Reflectivity_thin_metal.eps}(d). The curve for $r$ contains only one loop resulting in
the single dip in reflectance spectrum in Fig.\ \ref{Reflectivity_thin_metal.eps}(a).
The phase of the total reflection coefficient $r$ is $\pi$ in the middle of the resonance.
Due to the reduced splitting between the resonances the 
loops of $r_s$ and $r_a$ approach each other in Figs.\ \ref{Absorbance_modes.eps}(b)-(d), which leads to
exponential decay of the transmittance $T$ shown in Figs.\ \ref{Reflectivity_thin_metal.eps}(b)-(c).

In Fig.\ \ref{field_graphs.eps}  we show the field distribution during an oscillation period
for several situations. In panels (a)-(e) the magnetic field is shown for the first half of the period,
while electric field is shown for the second half. 
The graphs correspond to normal incidence, therefore
the field is not modulated perpendicularly to the growth axis. 
Off-resonance incidence upon the structure leads to full reflection of radiation, without considerable field intensity penetrating
the structure itself. If the incident radiation excites the symmetric field resonance
as shown in Fig.\ \ref{field_graphs.eps}(a), the field penetrates into the structure.
The character of the field distribution in Fig.\ \ref{field_graphs.eps}(a) 
confirms that the symmetric resonance is excited. 
The parameters pertaining to the panels of Fig.\ \ref{field_graphs.eps} are marked 
in Fig.\ \ref{Reflectivity_thin_metal.eps}(c). 

In Fig.\ \ref{field_graphs.eps}(b) excitation of an antisymmetric resonance is demonstrated. 
In this case the magnetic field 
is out of phase on different sides of the structure. One can also notice that there is a small delay of the field oscillations 
on the upper side of the structure with respect to the field at the bottom. This is due to admixture of the 
symmetric mode to mainly antisymmetric field distribution. 
At $n_L=3.1$ the dielectric contrast is already high enough for
the symmetric and antisymmetric modes to be spectrally close and interfere. High degree of interference is shown in
Fig.\ \ref{field_graphs.eps}(c) for excitation at a frequency between the symmetric and antisymmetric resonances
(see Fig.\ \ref{Reflectivity_thin_metal.eps}(c)). The field is enhanced at the bottom of the structure and suppressed
at the top, which corresponds to exponential growth of the field towards the bottom of the structure. 

Interference in the structure can be also seen in Fig.~\ref{field_graphs.eps}(d)
at the merger point of the symmetric and antisymmetric modes. Both modes are strongly excited and interfere to
produce field distributions which are localized on one side of the structure and decay towards another side
with the localization side being alternated after a quarter of the optical period.
After the full merger of the resonances the modes destructively interfere on the bottom side of the structure
as shown in Figs.\ \ref{field_graphs.eps}(e)-(f). The field decays exponentially towards the back side of the structure,
which results in exponential decay of transmission as was discussed in relation with Figs.\ \ref{Reflectivity_thin_metal.eps}(b)-(c).
The power that is not transmitted is dissipated in the top metal film instead, which 
is represented in Fig.\ \ref{Absorbance_modes.eps}(a) by increased absorbance when the splitting between 
the symmetric and antisymmetric resonances is small.

In summary, we have shown that optical properties of the
metal nanolayer - DBR - metal nanolayer structure in the frequency range near the stop band of the DBR
are determined by a pair of resonances which are intermediate between Fabry-P\'erot resonances
at low dielectric contrast of the DBR and a Tamm plasmon at high contrast. 
We have demonstrated that in a MNL-DBR-MNL structure with inversion symmetry
the low energy resonance corresponds to a resonance of field part, which is symmetric in magnetic field, 
while the antisymmetric part has a higher energy resonance. 
The spectral merger of these resonances leads to appearance of a Tamm plasmon and
considerable decrease of transmission as well as enhanced absorption.
In moderate and high contrast structures the resonances are weakly split, which allows for
various spatio-temporal interference patterns and opens possibility for coherent control in the proposed structure.

\end{document}